\documentclass[prc,twocolumn,english,showpacs]{revtex4}
\usepackage{graphics}

\makeatletter

\begin{document}

\title{Combined method to extract spectroscopic information}
\author{A.M. Mukhamedzhanov}\email{akram@comp.tamu.edu}
\affiliation{Cyclotron Institute, Texas A\& M University,
College Station, TX 77843, USA}
\author{F.M. Nunes}\email{nunes@nscl.msu.edu}
\affiliation{N.S.C.L. and Department of Physics and Astronomy, 
Michigan State University, U.S.A.}


\begin{abstract}
Spectroscopic factors (SF) play an important role in nuclear physics and 
astrophysics. 
The traditional method of extracting SF from 
direct transfer 
reactions suffers from serious ambiguities.
We discuss a modified method which is based on including the asymptotic 
normalization coefficient (ANC) of the overlap functions into 
the transfer analysis. 
In the modified method the contribution of the external part of the reaction 
amplitude, typically dominant, is fixed and
the SF is determined from fitting the internal part.  
We illustrate the modified method with $(d,p)$ reactions on 
${}^{208}{\rm Pb},\,\,{}^{12}{\rm C}$, and ${}^{84}{\rm Se}$ targets at 
different energies. 
The modified method allows one to extract the SF, which
do not depend on the shape of the single-particle nucleon-target 
interaction, and has the potential of improving the reliability 
and accuracy of the structure information.
This is specially important for nuclei on dripline, where not
much is known. 
\end{abstract}
\maketitle

SF were introduced by the shell model formalism 
and are typically related to the shell occupancy of a state $n$ in one nucleus 
relative to  a state $m$ in a nearby nucleus \cite{brown}. 
Today, phenomenological SF are extensively 
used in a variety of topics, from nuclear reactions to astrophysics or 
applied physics,  yet the procedure for
their extraction from the data has remained essentially the same for
decades. For more than forty years since the dawn of nuclear physics, direct 
transfer reactions, such as 
$(d,p),\,(d,t),\,({}^{3}{\rm He},d),\,({}^{3}{\rm He},\alpha)$,
have been the central tool to determine SFs 
\cite{goncharov,austern70,jpg-rev}.
Extracting SFs with good precision from data
is very important to test the validity of today's many body theories. 
For conventional nuclei there are many experiments available providing SFs,
which are often lower than those predicted by shell model \cite{brown}.   
Electron-induced knockout or electron scattering 
is supposed to provide a better accuracy in extracting SFs than transfer 
\cite{knock,kramer}. 
However, for exotic nuclei near or on the driplines,
transfer reactions are a unique tool and, hence, can have a large
impact in the programs of the new generation rare isotope laboratories.
Given the experimental difficulties faced with measurements 
on the driplines, it is crucial to have a reliable method for 
analyzing and extracting useful information from each single data set.

Usually, transfer angular distributions are
analyzed within the framework of the distorted-wave Born approximation 
(DWBA). The SF determined by normalizing the calculated DWBA 
differential cross section to the experimental one 
(e.g. \cite{schiffer,dwba,iliadis04}) is compared with the 
SF predicted by shell model.
Even when error bars in the experimental
cross section are low, the uncertainty of the extracted SF resulting
from the normalization of the DWBA cross section is often
large, regardless of whether it agrees with the shell model prediction.
The reasons for this inaccuracy are typically: i)  optical potentials 
ambiguity, ii) the inadequacy of the DWBA reaction theory, or iii) 
the dependence
on the single-particle potential parameters. 
The first point has been object of
a recent systematic study \cite{liu}. The second point needs to
be addressed case by case, and examples of improved reaction models are 
the coupled channel Born approximation (e.g. \cite{ccba})
or the continuum discretized coupled channel method (e.g. \cite{hirota}).
This work will critically review the standard procedure 
of extracting SFs from transfer reactions focusing 
on the third point; the modified method eliminates 
the dependence of the extracted SFs on the single-particle potentials,  
the main advantage of the method. 

We will address a modified approach to spectroscopy from transfer
reaction which includes the asymptotic normalization coefficient (ANC) 
in the analysis \cite{goncharov}. 
For simplicity, in the following formulation, we consider $A(d,p)B$ 
reaction and disregard spins (naturally these are included
in the applications). The DWBA amplitude for this
reaction is given by:
\begin{equation}
M= <\psi_{f}^{(-)}I^{B}_{An}|\Delta V|\varphi_{pn}\,\psi_{i}^{(+)}>,   
\label{dwba1}
\end{equation}  
where $\Delta V= V_{pn} + V_{pA} - U_{pB}$ is the transition 
operator in the post-form,
$V_{ij}$ is the interaction potential between $i$ and $j$, 
$U_{pB}$ is the optical potential in the final-state.
The distorted waves in the initial and final states are
$\psi^{(+)}_{i}$ and $\psi^{(-)}_{f}$,
$\,\varphi_{pn}$ is the deuteron bound-state wave function
and $I^{B}_{An}({\rm {\bf r}})$ is the overlap function of the 
bound-states of nuclei $B$ and $A$ which
depends on  $\,{\bf {\rm \bf r}}$, the radius-vector connecting the 
center of mass of $A$ with $n$. 
The overlap function is not an eigenfunction of an Hermitian Hamiltonian 
and is not normalized to unity \cite{blokh77}. 
The square norm of the overlap function gives a model-independent 
definition of the SF:
\begin{equation}
S= N\,<I^{B}_{An}|I^{B}_{An}>. 
\label{spectrfact1}
\end{equation}
Here, $N$ is the antisymmetrization factor in the isospin 
formalism (N will be included in the overlap function from now on). 

The leading asymptotic term of the radial overlap function (for $B= A +n$) is  
\begin{equation}  
I^{B}_{An(lj)}(r) \stackrel{r > R}{\approx} 
C_{lj}\,i\,\kappa\,h_{l}(i\,\kappa\,r),
\label{asympovfunct1}
\end{equation}
where $h_{l}(i\,\kappa\,r)$ 
is the spherical Bessel 
function, $\kappa= \sqrt{2\,\mu_{An}\,\varepsilon_{An}}$, 
$\varepsilon_{An}$ 
is the binding energy for $B \to A + n$, and $\,\mu_{An}$ 
is the reduced mass of $A$ and $n$.
Similarly, the asymptotics of the neutron single-particle wave function is
$\varphi_{An(n_{r}lj)}(r)\stackrel{r > R}{\approx}  
b_{n_{r}lj}\,i\,\kappa\,h_{l}(i\,\kappa\,r)$, where 
$n_{r}$ is the principle quantum number. 
The asymptotic behaviour is valid beyond $R$, the channel radius.
It is clear that, in the asymptotic region, the 
overlap function is proportional to the single particle wave function.
The normalization $C_{lj}$ introduced in Eq.(\ref{asympovfunct1})  
is the ANC which relates to the single-particle ANC (SPANC) $b_{n_{r}lj}$ by 
$C_{lj}= K_{n_{r}lj}\,b_{n_{r}lj}$, 
where $K_{n_{r}lj}$ is an asymptotic proportionality coefficient. 
It is standard practice to assume that the proportionality between
the overlap function and the single particle function extends 
to all r values
\begin{equation}
I^{B}_{An(lj)}(r)= K_{n_{r}lj}\,\varphi_{An(n_{r}lj)}(r).
\label{overlap}
\end{equation}
Since  $\varphi_{An(n_{r}lj)}(r)$ is normalized to unity, this approximation
(Eq. \ref{overlap}) implies that  $S_{lj}= K^{2}_{n_{r}lj}$. 
We have to emphasize, however, that the overlap function 
in the interior is nontrivial and may well differ  from
the single particle wavefunction. Approximating the radial dependence
of the overlap function as described above leads to the
DWBA amplitude 
\mbox{$M= K_{n_{r}lj}<\psi_{f}^{(-)}\varphi_{An(n_{r}lj)}|\Delta
V|\varphi_{pn}\,\psi_{i}^{(+)}>$.}
Normalizing the  calculated  DWBA cross section,
\begin{equation}
\sigma^{DW}=   
|<\varphi_{An(n_{r}lj)}|\Delta V|\varphi_{pn}\,\psi_{i}^{(+)}>|^{2}
\label{dwbasp}
\end{equation}
to the experimental data provides the phenomenological  SF $S_{lj}=K_{n_{r}lj}^{2}$.
Assuming that Eq.(\ref{overlap}) is  valid for all r,
we can infer from Eq.(\ref{spectrfact1}) that the main contribution to the 
norm of the overlap function comes from the nuclear interior. 

In order to make the dependence on the SPANC more explicit, 
we split the reaction amplitude 
into an interior part and an exterior part:
\begin{eqnarray}
M&=& K_{n_{r}lj}\,{\tilde M}_{int}[b] +  
K_{n_{r}lj}\,b_{n_{r}lj}\,{\tilde M}_{ext},                                
\label{reactam1}
\end{eqnarray}
where the internal part of the matrix element
\mbox{${\tilde M}_{int}[b_{n_{r}lj}] = <\psi_{f}^{(-)}\varphi_{An(n_{r}lj)}|\Delta
V|\varphi_{pn}\,\psi_{i}^{(+)}>_{r < R}$}
depends on $b_{n_{r}lj}$
through the bound state wavefunction $\varphi_{An(n_{r}lj)}$,
while the external part
${\tilde M}_{ext}= \,<\psi_{f}^{(-)}\,i\,\kappa\,h_{l}(i\,\kappa\,r)|\Delta V|
\varphi_{pn}\,\psi_{i}^{(+)}>_{r > R}$  does not depend on 
$b_{n_{r}lj}$. Here,
$R$ is the channel radius taken so that for $r > R$ the overlap 
function can be approximated by its asymptotic form Eq.(\ref{asympovfunct1})
(R is only used to illustrate the method as in the end
this separation is not required).
The contribution from the nuclear exterior is fixed by the ANC,
whereas the SF determines the normalization of the internal 
part of the radial matrix element. Since transfer reactions are dominantly peripheral,
SFs can only be extracted from transfer reactions due to a 
small contribution from the nuclear interior. 
We now introduce the ANC into the DWBA cross section:
\begin{equation}
\frac{ {\rm d}\,\sigma^{DW} }{{\rm d}\Omega}=
C_{lj}^{2}\,\frac{\sigma^{DW}}{b_{n_{r}lj}^{2}}.
\label{dwbaxs}
\end{equation}
Introducing Eq.(\ref{reactam1}) into Eq.(\ref{dwbaxs}) 
and dividing by $C_{lj}^{2}$,
we arrive at a function $R^{DW}(b)$
\begin{equation}
R^{DW}(b_{n_{r}lj})= |\frac{{\tilde M}_{int}[b]}{b_{n_{r}lj}}+ 
{\tilde M}_{ext}|^{2}.
\label{ratio}
\end{equation}
Note that the single-particle ANC  $b_{n_{r}lj}$ itself is a function of the 
geometrical parameters of the bound state $n-A$ nuclear potential $(r_0,a)$
which are, a priori, not known. 
If the  ANC and the cross section for the (d,p) reaction 
have been measured, the experimental counterpart of $R^{DW}$,
$R^{exp}= \frac{{\rm d}\,\sigma^{exp}}{{\rm d}\Omega}/C_{lj}^{2} $ 
can be experimentally fixed.    
Then, imposing the equality
\begin{eqnarray}
R^{exp} = R^{DW}(b_{n_{r}lj}),
\label{dwrfunct1}
\end{eqnarray}
will provide the correct $b_{nlj}$ 
and consequently the SF $S_{lj}=C_{lj}^2/b_{nlj}^2$.

At this stage, a few points should be made clear. 
First of all, for specific optical potentials, Eq. (\ref{dwbasp}) 
depends on two independent parameters, 
$S_{lj}$ and $b_{nlj}$. In the standard approach, to evaluate this cross section, 
the second parameter is fixed by arbitrarily
choosing the bound state $n-A$ potential geometry. 
Thus, the extracted product $S_{lj}\,b_{n_{r}lj}^{2}$ does not 
coincide necessarily with the correct ANC. 
Since the ANC determines the normalization of the external part of the 
DWBA amplitude, in the standard approach
the SF is determined by an unrealistic variation of the external contribution. 
In the modified method here discussed, since the 
contribution of the external part is fixed through the correct ANC,
the whole DWBA  procedure loses this artificial degree of freedom. 

Secondly, if the reaction is peripheral, 
i. e. the first term in Eq. (\ref{reactam1}) is negligible, 
one can determine the ANC. So, the modified approach makes use of two experiments:
the first to fix the ANC, the second to determine the SF consistent with
that ANC.
In present experiments and with the new generation of rare isotope facilities, 
ANCs can be determined with $5$\% accuracy. 
Since the determination of the SF comes from the internal region,
the second experiment needs to be performed at a beam energy 
for which the contribution from the interior is significant. 
The higher the contribution of the internal region, 
the stronger the dependence on $b_{n_{r}lj}$ in $R^{DW}(b_{n_{r}lj})$ 
and the smaller the uncertainty of the extracted SF, although 
a balance needs to be found since large interior
contributions may not be well describe by DWBA.
The DWBA differential cross section near the main peak of 
the angular distribution and, correspondingly, $R^{DW}(b_{n_{r}lj})$ 
are the functionals 
of the single-particle ANC $b_{n_{r}lj}$. 
One given $b_{n_{r}lj}$ can be produced by an infinite number 
of single-particle potentials, local and non-local. However, 
the dependence of  ${\rm d}\,\sigma^{DW}/{\rm d}\Omega$ or
$R^{DW}(b_{n_{r}lj})$ on the shape of the single-particle potential 
is minor.
Hence, the extracted SF in the modified method does not depend on 
the single-particle potential. 
We illustrate the method presenting three different applications: 
i) $^{209}$Pb, ii) $^{13}$C and iii) $^{85}$Se. We will drop the subscripts
on $b$ for simplicity.

\begin{figure}[t]
\resizebox*{0.3\textwidth}{!}{\includegraphics{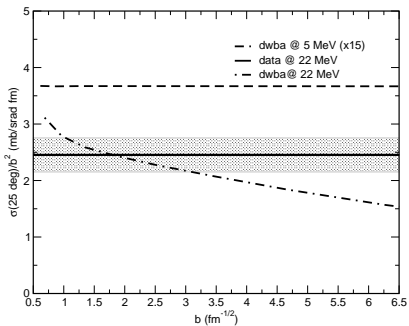}}
\vspace{-0.2cm}
\caption{\label{pb208dp} Cross section for $^{208}$Pb(d,p)$^{209}$Pb(g.s.)
at 22 MeV and the dependence on the single particle parameters:
experimental value (solid line), experimental error bar (shaded area)
and the DWBA prediction (dot-dashed).}
\vspace{-0.5cm}
\end{figure}
Let us consider the reaction $^{208}$Pb(d,p)$^{209}$Pb from
\cite{hirota}. Although the ANC for $<^{209}$Pb$|^{208}$Pb$>$ is not 
published, it can be determined from the sub-Coulomb reaction \cite{franey}
$^{208}$Pb($^{13}$C,$^{12}$C)$^{209}$Pb as the other vertex 
$<^{13}C|^{12}C>$ is well known \cite{c13anc}.
Sub-Coulomb reactions are extremely peripheral and insensitive to
details of the optical potentials. For this reason they present an
excellent probe for extracting the ANC accurately.
From \cite{franey} we obtain an ANC $C^2_{g9/2}=2.15(0.16) $fm$^{-1}$
for $^{209}$Pb. Then using $^{208}$Pb(d,p)$^{209}$Pb 
data at $E_d=22$ MeV \cite{hirota} we obtain 
$R^{exp}=2.46(0.31)$ fm mb/srad, where the error bar is calculated
based on both, the ANC and the cross sections errors,  taken
as independent.
The experimental data in \cite{hirota} has $1$\% accuracy but is taken
only down to $\theta_{cm} = 35 ^{\circ}$ 
whereas the peak of the DWBA distribution
is at $\theta_{cm}=25 ^{\circ}$. 
We extrapolate the data based on the shape predicted
by DWBA and include a 10\% error in the cross section to account for this 
difference. Measurements at $25 ^{\circ}$ could improve the error bar
in $R^{exp}$ considerably.
We next perform a series of finite range DWBA calculations
for $^{208}$Pb(d,p)$^{209}$Pb ($E_d=22$ MeV), 
using the optical potentials from \cite{perey}. The adiabatic prescription
\cite{soper} was used to take into account deuteron breakup
which is important for this reaction. The Reid-soft-core
potential was used for the deuteron wavefunction, as well
as in all other examples. 
For illustration purposes, we use a Woods Saxon well to generate the $^{208}Pb+n$ 
single-particle wavefunctions and obtain a range of SPANCs $b$ by
varying the single particle parameters $(r_0,a)$ and adjusting 
the depth to reproduce the correct binding for
the $2g_{9/2}$ in each case. We use the same s.o. 
strength as that in \cite{franey} although the
s.o. strength does not affect the final result.

The results of our calculations $R^{DW}$ (dot-dashed line) and 
the experimental value 
$R^{exp}$ (solid line and shaded area) are presented, 
as a function of b, in Fig. \ref{pb208dp}. From $R^{exp}$ 
one finds $b=1.82 $ fm$^{-1/2}$ and S=0.74.
It is worth noting that in the standard approach typical parameters 
$(r_0,a)=(1.2,0.6)$ fm,  produce $b=1.34$ fm$^{-1/2}$. 
The direct comparison of the DWBA cross section using $(r_0,a)=(1.2,0.6)$ fm,
with the data, give S=0.866 and consequently,  $C^{2}=1.56$\,fm$^{-1}$, 
beyond the experimental range. 
As pointed out before, in the standard approach the SF is determined 
at the cost of an artificial ANC. 

The beam energy of 22 MeV is above the Coulomb
barrier, thus the reaction is not peripheral. This can be seen in
Fig. \ref{pb208dp} through the slope of the dot-dashed curve.
In fact for this particular energy, the interior contribution is
around $10$\%. The uncertainty  in $b \in [1.1,3.1]$ fm$^{-1/2}$ propagates
into a large uncertainty in $S \in [0.3,2.2]$. 
This is due to the fact
that the contribution from the interior at this energy is still small.
The scaling factor relating the uncertainty of S with that of b
is $(|M_>|/|M_<|)^2$. The smaller the contribution from the interior,
the smaller the accuracy with which the SF can be determined.

Also in Fig. \ref{pb208dp} we show the results for $R^{DW}$ 
corresponding to the calculation at $E_d=5$ MeV (dashed line). This is to 
illustrate that, at sub-Coulomb energies,
the reaction becomes completely peripheral and the dependence
on $b$ disappears. Measurements at these energies could provide
$C^2_{g_{9/2}}$ with accuracy $<5$\%. In addition, measurements at
higher energy ($> 30$ MeV) would increase the slope of $R^{DW}(b)$ and
decrease further the error on the extracted SF.

\begin{figure}[t]
\resizebox*{0.3\textwidth}{!}{\includegraphics{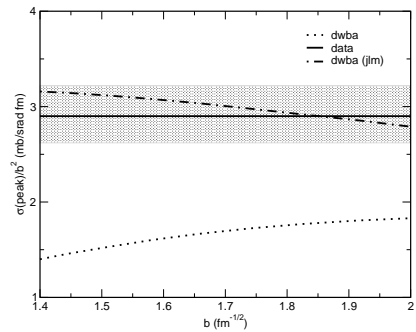}}
\vspace{-0.2cm}
\caption{\label{c12dp} Cross section for $^{12}$C(d,p)$^{13}$C(g.s.)
at 51 MeV and the dependence on the single particle parameters:
experimental value (solid line), experimental error bar (shaded area)
and the DWBA prediction (dot-dashed).}
\vspace{-0.5cm}
\end{figure}
Another standard case is the $^{12}$C(d,p)$^{13}$C reaction, for which
many data sets are conveniently compiled in a recent publication 
\cite{liu}. We studied three cases (8.9 MeV, 30 MeV and 51 MeV),
using the same JLM optical potentials as \cite{liu}. We 
perform a series of finite range DWBA calculations 
varying the $1p_{1/2}$ $^{12}$C-n single particle parameters,
in order to obtain 
$R^{DW}(b)$ as described before. Results 
for the less peripheral case (51 MeV) are plotted in Fig. \ref{c12dp}
(dotted-dashed line).
We take the data from \cite{liu} and the ANC from \cite{c13anc},
to obtain 
$R^{exp}=\frac{\sigma(2.5 ^\circ)}{C^2_{1,1/2}}=2.92(0.35)$ fm mb/srad. 
An $S=0.66$ (shell model) would require $b=1.89$ which
is contained in our results. However, such a conclusion is misleading.
Fig. \ref{c12dp} shows that even for this relatively 
large energy, the dependence of $R^{DW}$ on $b$ is weak.
Consequently, it is not possible to extract a SF. 

It was pointed out in \cite{liu} that the deuteron breakup
is important for this reaction and should be taken into account. 
To emphasize this fact, we compare our results using
the adiabatic deuteron potential \cite{soper} from \cite{liu} 
(dot-dashed line in Fig. \ref{c12dp})
with those obtained using an optical potential fitted to the deuteron
elastic scattering (dotted line in Fig. \ref{c12dp}). The disagreement
is very large. Interestingly, the method here described is also
able to detect inadequate optical potential parameterizations.

Oak-Ridge has developed a program
to measure a series of inverse kinematics (d,p) reactions 
for nuclei on the neutron dripline \cite{oak}. As one of the nuclei
in the program is $^{85}$Se,
we have performed exploratory calculations for 
$^{84}$Se(d,p)$^{85}$Se. We take global parameterizations
for the optical potentials \cite{perey} and perform a series
of calculations varying the single particle parameters. We 
compare the dependence of $R^{DW}$ on $b$ for a range of energies
$E_d=4-100$ MeV.
We verify that, expectedly,
the dependence on $b$ increases with beam energy. We find that 
Oak-Ridge energies (10 MeV/A) are adequate to determine ANCs 
but not SFs. However, a facility that allows for
the production of $^{84}$Se at $E>25$ MeV/A  (such as NSCL-MSU, GANIL or RIKEN) 
could provide accurate spectroscopic information.

In conclusion, we have presented an alternative method to extract
SFs, taking into account the sensitivity
of the transfer data to the interior part of the overlap function
and combining that information with the ANC.
Transfer data can only become useful within this
method if it has a significant contribution from the interior, and is well
described through a one-step DWBA formalism. The balance between 
these two conditions is not a trivial one.  
By reducing the error bars in both the measured transfer cross section
and the ANC, this prescription
determines the single particle asymptotics and from it, a 
SF with reduced uncertainty. 
The ANC needs to be determined independently;
it can be pinned down accurately with
the same transfer reaction at sub-Coulomb energies or 
using heavy-ion induced reactions, both safely peripheral.
Note that uncertainties due to optical potentials and 
higher order effects need to be assessed independently, as this
work focuses on the single particle parameter uncertainties only.

The method here presented has the potential of reducing 
the uncertainty in the overlap function considerably. 
However it still assumes that the 
interior part has a Woods Saxon single particle 
wavefunction shape. This has been corroborated by
recent Green's Function Monte Carlo calculations on light nuclei
\cite{wiringa}.
Even if there were non-localities of the single particle potential 
this would affect mostly the deep interior and thus would not be
visible in the transfer reactions.   

Results for (d,p) on $^{208}$Pb were used to illustrate the method.
We discussed previous analyzes of  (d,p) reactions  on 
$^{12}$C, and showed the limitations. We have also demonstrated 
that this method can rule out inadequate choices of optical potentials. 
Considering specific future experiments, we have performed
exploratory calculations for (d,p) on $^{84}$Se. 
This method will become useful for a broad variety of transfer experiments
in the field of rare isotopes. The same method can equally be used for transfer
to excited states. These same ideas can be extended to other
reactions, in particular breakup reactions which also
have an impact on Astrophysics.
Finally, it would be helpful if the state-of-the-art reaction codes
would incorporate the formalism discussed.

\small
We thank Xiandong Liu for providing the JLM potentials
and Prof. Goldberg for useful comments.
This work has been partially
supported by the NSCL at Michigan State
University, U.\ S.\ DOE under Grant No.\
DE-FG03-93ER40773, by NSF Award No.\ PHY-0140343 and by 
Funda\c{c}\~ao para a Ci\^encia e a Tecnologia (F.C.T.)
of Portugal, under the grant POCTIC/36282/99.

\end{document}